\documentclass[a4paper,10pt]{article}
\usepackage{latexsym}
\usepackage{amsmath}
\usepackage{amssymb}
\usepackage{graphicx}
\usepackage{wrapfig}
\usepackage{fancybox}
\usepackage{bm}
\title{The effect of an imaginary part of the Schwinger-Dyson equation at finite temperature and density}
\author{S.Sasagawa and H.Tanaka \\\\Department of Physics, Rikkyo University, Tokyo 171-8501, Japan}
\date{}
\usepackage{fancyhdr}
\makeatletter
\def\firstpage{\hfill RUP-11-6}
\def\ps@titlepage{%
   \@oddhead{\hfil\firstpage\hfil}%
   \let\@evenhead\@oddhead
   \def\@oddfoot{\hfil}%
   \let\@evenfoot\@oddfoot
   \let\@mkboth\@gobbletwo}
\makeatother

\begin{document}
\vspace{1em}
\maketitle

\thispagestyle{titlepage}

\vspace{1em}
\begin{abstract}

\noindent We examined the effect of an imaginary part of the ladder approximation Schwinger-Dyson equation. We show the imaginary part enhances the effect of the first order transition, and affects a tricritical point. In particular, a chemical potential at a tricritical point is moved about $200$ (MeV). Thus, one should not ignore the imaginary part. On the other hand, since an imaginary part is small away from a tricritical point, one should be able to ignore an imaginary part. In addition, we also examined the contribution of the wave function renormalization constant.

\end{abstract}

\vspace{1em}
$\\[0.28cm]$

\vspace{1em}
\vspace{1em}
\section{Introduction}

The chiral symmetry breaking is restored at high temperature and density. The chiral symmetry breaking and restoration at zero and finite temperature are extensively studied by the lattice QCD simulation. The quark-gluon plasma at high temperature has been observed at the Relativistic Heavy Ion Collider \cite{rf:ex1}. On the other hand, there is an incomplete understanding of phenomena at low temperature and high density. Although it is expected that various phases exist in the region of low temperature and high density \cite{rf:dense1}, we do not have sufficient information. For example, the lattice QCD at low temperature and high density is still inadequate by the sign problem. (This problem arises from the fermion determinant that has complex. Several methods for the sign problem were developed, see e.g., \cite{rf:la2,rf:la3,rf:la4}.) Furthermore, in addition to the incomplete lattice simulation, there is little information by experimental data. Thus, it is very important to investigate this region.

The Schwinger-Dyson equation (SDE) is an useful tool at non-zero temperature and chemical potential \cite{rf:1,rf:2,rf:3,rf:4,rf:5}. This method can be used at low temperature and high density. For example, the SDE with the hard dense loop approximation can derive the mass gap for color superconductivity \cite{rf:5}. For the chiral symmetry breaking and restoration, the SDE shows that the phase transition is of second order at finite temperature, and is of first order at non-zero chemical potential \cite{rf:1,rf:2,rf:4}. This has been shown in the context of models for the spontaneous broken chiral symmetry (other than QCD), i.e., the Nambu-Jona-Lasinio model \cite{rf:Buba}.

The SDE at non-zero chemical potential has an imaginary part. The imaginary part is generated by the presence of a chemical potential. Thus, the imaginary part has the effect of a chemical potential. If the imaginary part does not affect the phase transition strongly, one can ignore it to simplify the calculation. However, this has been studied in \cite{rf:4}, and it was shown that the effect of the imaginary part is strong.

Since the approximate form for the exact quark propagator was used in \cite{rf:4}, we verify the effect of the imaginary part using the more general form for the exact quark propagator at finite temperature and density. (We do not consider the Debye screening and the 2 flavor superconductivity.) The more genaral quark propagator at non-zero temerature and density is the $SO(3)$ invariant form, and has the wave function renormalization constant. We also study this effect for a tricritical point.

The paper is organized as follows. In section\ \ref{sec:im} and \ref{sec:sd}, we review the formulation at non-zero temperature and chemical potential. We show numerical results by solving the SDE in section\ \ref{sec:num}, and a summary is found in section\ \ref{sec:sum}.

\vspace{1em}
\section{Formulation at non-zero temperature and chemical potential}\label{sec:im} 

We use the imaginary time formalism to analyze at non-zero temperature and chemical potential \cite{rf:fetter,rf:bellac}. The ensemble average of an operator at temperature $ T=1/\beta$ and chemical potential $\mu$ is defined as

\begin{equation}
\langle\hat{O}\rangle_{\beta}=Z^{-1}\mathrm{tr}[e^{-\beta(\hat{H}-\mu\hat{Q})}\hat{O}],\ Z=\mathrm{t}\mathrm{r}[e^{-\beta(\hat{H}-\mu\hat{Q})}],
\end{equation}

\noindent where $\hat{H}$ and $\hat{Q}$ are a Hamiltonian and a number operator. The partition function $Z$ can be rewritten by

\begin{equation}
Z=\displaystyle \int \mathcal{D}\phi \mathcal{D}\phi^{*}\int \mathcal{D}\pi \mathcal{D}\pi^{*}\exp\left[\int_{0}^{\beta}d\tau\int d^{3}x\left(i\pi\frac{\partial\phi}{\partial\tau}+i\pi^{*}\frac{\partial\phi^{*}}{\partial\tau}-\mathcal{H}(\pi,\pi^{*},\phi,\phi^{*})+\mu Q\right)\right],
\end{equation}

\noindent where $\tau$ is an imaginary time $\tau=it,\ 0\leq\tau\leq\beta$. The field $\phi(\tau,\bm{x})$ has the periodicity $\phi(0,\bm{x})=\phi(\beta,\bm{x})$. When the Lagrangian of the complex scalar field is used, the partition function is

\begin{equation}
Z=N(\displaystyle \beta)\int \mathcal{D}\phi \mathcal{D}\phi^{*}\exp\Big[\int_{0}^{\beta}d\tau\int d^{3}x\mathcal{L}\Big],\label{eq:par}
\end{equation}

\noindent with

\begin{equation}
\displaystyle \mathcal{L}=-(\frac{\partial\phi}{\partial\tau}-\mu\phi)(\frac{\partial\phi^{*}}{\partial\tau}+\mu\phi^{*})-|\nabla\phi|^{2}-m^{2}|\phi|^{2}
\end{equation}

\noindent This form is the same as the Euclidean functional integral in the field theory. Thus, we can use the same approach as the zero temperature field theory. 

After performing the $\pi$ integral, the partition function (\ref{eq:par}) is 

\begin{equation}
Z=\displaystyle \int \mathcal{D}\phi \mathcal{D}\phi^{*}\exp\Big[\int_{0}^{\beta}d\tau\int d^{3}x\phi^{*}((\frac{\partial}{\partial\tau}-\mu)^{2}+\nabla^{2}-m^{2})\phi\Big]
\end{equation}

\noindent Owing to the periodicity that $\phi(0,\bm{x})=\phi(\beta,\bm{x})$, the Fourier transformation is

\begin{equation}
\displaystyle \phi(\tau,\bm{x})=T\sum_{n=-\infty}^{+\infty}\int\frac{d^{3}p}{(2\pi)^{3}}e^{-i(\omega_{n}\tau-\bm{p}\cdot \bm{x})}\phi_{n}(\bm{p})
\end{equation}

\noindent where $\omega_{n}$ is the Matsubara frequency, $\omega_{n}=2n\pi T (n=0,\pm 1,\pm 2,\ldots)$ for bosons. Using this Fourier transformation, from analogy with zero temperature field theory, the free propagator $D_{\beta}(i\omega_{n},\bm{p})$ is given by

\begin{equation}
D_{\beta}(i\displaystyle \omega_{n},\bm{p})=\frac{-1}{(i\omega_{n}+\mu)^{2}-|\bm{p}|^{2}-m^{2}}
\end{equation}

\noindent To distinguish between a imaginary time green function and a real time green function, we refer to $D_{\beta}(i\omega_{n},\bm{p})$ as the thermal green function. The thermal green function is obtained from the Minkowski propagator by the replacement,

\begin{equation}
\displaystyle \frac{i}{p^{2}-m^{2}}\ \Rightarrow\ \frac{-1}{p^{2}-m^{2}},\ p_{0}=i\omega_{n}+\mu.\label{eq:rep}
\end{equation}

Similarly, the partition function for fermions is given by

\begin{equation}
Z=\displaystyle \int \mathcal{D}\overline{\psi}\mathcal{D}\psi\exp\left[\int_{0}^{\beta}d\tau\int d^{3}x\ \left(\overline{\psi}(-\gamma_{0}\frac{\partial}{\partial\tau}+i\bm{\gamma}\cdot\nabla-m+\gamma_{0}\mu)\psi\right)\right].
\end{equation}

\noindent Since a fermion field has the antiperiodicity $\psi(0,\bm{x})=-\psi(\beta,\bm{x})$, the Fourier transformation for fermions is

\begin{equation}
\displaystyle \psi(\tau,\bm{x})=T\sum_{n=-\infty}^{+\infty}\int\frac{d^{3}p}{(2\pi)^{3}}e^{-i\omega_{n}\tau+i\bm{p}\cdot \bm{x}}\psi_{n}(\bm{p}).
\end{equation}

\noindent The Matsubara frequency for fermions is $\omega_{n}=2\pi T(n+1/2) (n=0,\pm 1,\pm 2,\ldots)$. Thus, the free thermal green function is

\begin{equation}
S_{\beta}(i\displaystyle \omega_{n},\bm{p})=\frac{-1}{(i\omega_{n}+\mu)\gamma_{0}-\bm{\gamma}\cdot \bm{p}-m}
\end{equation}

\noindent We can use the same replacement (\ref{eq:rep}) for fermions.

\vspace{1em}
\section{Schwinger-Dyson equation at non-zero temperature and chemical potential}\label{sec:sd} 

The partition function at QED is

\begin{equation}
Z=\displaystyle \int \mathcal{D}A_{\mu}\mathcal{D}\overline{\psi}\mathcal{D}\psi\exp\Big[\int_{0}^{\beta}d\tau\int d^{3}x\mathcal{L}\Big],
\end{equation}

\noindent with

\begin{equation}
\displaystyle \mathcal{L}=\overline{\psi}(-\gamma_{0}\frac{\partial}{\partial\tau}+\gamma_{0}\mu+i\bm{\gamma}\cdot\nabla-m-eA\hspace{-.50em}/)\psi-\frac{1}{4}F_{\mu\nu}F^{\mu\nu}-\frac{1}{2\xi}(\partial^{\mu}A_{\mu})^{2},
\end{equation}

\noindent Here, we used the Minkowski notation, $\partial_{\mu}=(i\partial/\partial\tau,\nabla)$ and $A_{\mu}=(A_{0},-\bm{A})$. Since the argument at QCD is essentially identical to QED, we use QED for simplicity. Adding sources, the generating functional is

\begin{equation}
Z[J_{\mu},\displaystyle \eta,\overline{\eta}]=\int \mathcal{D}A_{\mu}\mathcal{D}\overline{\psi}\mathcal{D}\psi\exp\Big[\int_{0}^{\beta}d\tau\int d^{3}x(\mathcal{L}+J_{\mu}A^{\mu}+\overline{\eta}\psi+\overline{\psi}\eta)\Big].
\end{equation}

\noindent This form is the same as the Euclidean generating functional apart from the integral range of $\tau$. Thus, using the procedure at the Euclidean or Minkowski space \cite{rf:robert}, the SDE for the fermion thermal green function is given by

\begin{equation}
G_{\beta}^{-1}(p)=-S_{\beta}^{-1}(p)-e^{2}T\displaystyle \sum_{l}\int\frac{d^{3}k}{(2\pi)^{3}}\gamma_{\mu}D_{\beta}^{\mu\nu}(k)G_{\beta}(p-k)\Gamma_{\nu}(p,k),\label{eq:sde}
\end{equation}

\noindent where $G_{\beta}$ is the exact fermion thermal green function, $D_{\beta}^{\mu\nu}$ is the exact photon thermal green function, and $\Gamma_{\nu}$ is the vertex function. $ p_{0}=i\omega_{n}+\mu$ is the Matsubara frequency for fermions and $k_{0}=i\omega_{l}$ is the Matsubara frequency for bosons. The free photon thermal green function is obtained by the replacement (\ref{eq:rep}),

\begin{equation}
\displaystyle \frac{1}{k^{2}}\Big(g^{\mu\nu}+(\xi-1)\frac{k^{\mu}k^{\nu}}{k^{2}}\Big).
\end{equation}

\noindent Since the thermal green function has a imaginary time or the discrete Matsubara frequency, the thermal green function is not directly physical quantity. Owing to this, it is different to the zero temperature SDE, and the SDE in the imaginary time formalism should be incomplete to study the chiral phase transition.

To study the chiral phase transition, we use the Cornwall-Jackiw-Tomboulis (CJT) effective potential \cite{rf:cjt}. The CJT effective potential at QED is given by

\begin{equation}
V[G]=-T\displaystyle \sum_{n}\mathrm{tr}\int\frac{d^{3}p}{(2\pi)^{3}}\log[G_{\beta}^{-1}(p)S_{\beta}(p)]-T\sum_{n}\mathrm{t}\mathrm{r}\int\frac{d^{3}p}{(2\pi)^{3}}S_{\beta}^{-1}(p)G_{\beta}(p)+V_{2}[G],\label{eq:cjt1}
\end{equation}

\noindent where

\begin{equation}
V_{2}[G]=\displaystyle \frac{e^{2}T^{2}}{2}\mathrm{tr}\Big[\sum_{n,m}\int\frac{d^{3}p}{(2\pi)^{3}}\int\frac{d^{3}q}{(2\pi)^{3}}\gamma_{\mu}G_{\beta}(p)D_{\beta}^{\mu\nu}(p-q)\gamma_{\nu}G_{\beta}(q)\Big]\label{eq:cjt2}.
\end{equation}

\noindent$\mathrm{tr}$ indicates a trace over spinor components. We eliminated irrelevant terms.

The exact fermion thermal green function can be written by

\begin{equation}
G_{\beta}(p)=\displaystyle \frac{-1}{C_{n}(\bm{p})\gamma_{0}p_{0}+A_{n}(\bm{p})\gamma_{i}p^{i}-B_{n}(\bm{p})},\label{eq:pro1}
\end{equation}

\noindent where $C_{n}(\bm{p}),\ A_{n}(\bm{p})$ and $B_{n}(\bm{p})$ are arbitrary scalar functions. This is the $SO(3)$ invariance form \cite{rf:2}. At zero temperature, $C(p)=A(p)$ are the wave function renormalization constant, and $B(p)$ is the mass function. Inserting (\ref{eq:pro1}) into (\ref{eq:sde}), the ladder approximation SDE \cite{rf:1,rf:2,rf:3,rf:robert,rf:ss} can be written by

\begin{equation}
C_{n}(x)=1+\displaystyle \frac{e^{2}T}{8\pi^{2}p_{0}x}\sum_{m}\int_{0}^{\infty}dyy\frac{-C_{m}(y)(I_{1}+I_{2})-A_{m}(y)I_{3}}{C_{m}^{2}(y)q_{0}^{2}-A_{m}^{2}(y)y^{2}-B_{m}^{2}(y)},\label{eq:gene1}
\end{equation}

\begin{equation}\\[0.35cm]
A_{n}(x)=1-\displaystyle \frac{e^{2}T}{8\pi^{2}x^{3}}\sum_{m}\int_{0}^{\infty}dyy\frac{-C_{m}(y)H_{1}+A_{m}(y)(H_{2}-H_{3})}{C_{m}^{2}(y)q_{0}^{2}-A_{m}^{2}(y)y^{2}-B_{m}^{2}(y)},\label{eq:gene2}
\end{equation}

\begin{equation}
\displaystyle \begin{split}B_{n}(x)=-T\frac{3e^{2}}{8\pi^{2}x}\sum_{m}\int_{0}^{\infty}dy&\displaystyle \frac{yB_{m}(y)}{C_{m}^{2}(y)q_{0}^{2}-A_{m}^{2}(y)y^{2}-B_{m}^{2}(y)}\\&\hspace{3.5em}\displaystyle \times\log\frac{(p_{0}-q_{0})^{2}-(x+y)^{2}}{(p_{0}-q_{0})^{2}-(x-y)^{2}},\end{split}\label{eq:gene3}
\end{equation}

\noindent where $x=|\mbox{\boldmath $p$}|,\ y=|\mbox{\boldmath $q$}|,\ p_{0}$ and $q_{0}$ are the Matsubara frequency for fermions, $I $and $H$ are shown in appendix$\ $A. The fermion is massless, and we adopted the Landau gauge $\xi=0$. Since $C_{n}(x),\ A_{n}(x)$ and $B_{n}(x)$ are complex functions, the SDE at non-zero chemical potential is constructed by six simultaneous equations (see appendix$\ $A).

Inserting (\ref{eq:pro1}) into (\ref{eq:cjt1}) and (\ref{eq:cjt2}), the ladder approximation CJT effective potential is

\begin{equation}
V=-2T\displaystyle \sum_{n}\int\frac{d^{3}p}{(2\pi)^{3}}\Big(\log[-C_{n}^{2}(\bm{p})p_{0}^{2}+A_{n}^{2}(\bm{p})|\bm{p}|^{2}+B_{n}^{2}(\bm{p})]+\frac{C_{n}(\bm{p})p_{0}^{2}-A_{n}(\bm{p})|\bm{p}|^{2}}{C_{n}^{2}(\bm{p})p_{0}^{2}-A_{n}^{2}(\bm{p})|\bm{p}|^{2}-B_{n}^{2}(\bm{p})}\Big).\label{eq:cjt3}
\end{equation}

\vspace{1em}
\noindent Since (\ref{eq:gene1})--(\ref{eq:gene3}) have the relation, $C_{n}(x)=C_{-n-1}^{*}(x), A_{n}(x)=A_{-n-1}^{*}(x)$, and $B_{n}(x)=B_{-n-1}^{*}(x)$, using this relation and $\omega_{n}=-\omega_{-n-1}$ for the fermion Matsubara frequency, one finds that ${\rm Im} V$ vanisehes. We can know a effect of the imaginary part of the SDE for the chiral phase transition by using this CJT effective potential.

We do not take into account of the Debye screening effect in the gluon thermal green function, because we guess that this contribution is unrelated to their imaginary parts. (The Debye screening effect by the hard thermal/dense loop approximation is a real \cite{rf:bellac}).

\vspace{1em}
\vspace{1em}
\section{Numerical calculation}\label{sec:num}

At QCD, the SDEs (\ref{eq:gene1})--(\ref{eq:gene3}) are replaced $e^{2}$ with $C_{2}g^{2}.\ C_{2}$ is the Casimir operator. Moreover, the coupling constant $g^{2}$ is replaced by a running coupling constant $g^{2}(-p^{2},-q^{2})$ (the improved ladder approximation). We adopt the form of a running coupling constant \cite{rf:1,rf:2};

\begin{equation}
g^{2}(-p^{2},-q^{2})=\beta_{0}\times\left\{\begin{array}{l}
\frac{1}{t}\ \ \hspace{8.8em},\ \ t_{F}<t,\\
\frac{1}{t_{F}}+\frac{(t_{F}-t_{C})^{2}-(t-t_{C})^{2}}{2t_{F}^{2}(t_{F}-t_{C})}\ \ \ ,\ \ t_{C}<t<t_{F},\\
\frac{1}{t_{F}}+\frac{t_{F}-t_{C}}{2t_{F}^{2}}\ \ \hspace{4.6em},\ \ t<t_{C},
\end{array}\right.\label{eq:Cou}
\end{equation}

\[
t=\log[(-p^{2}-q^{2})/\Lambda_{qcd}^{2}],\ \beta_{0}=\frac{48\pi^{2}}{11N_{c}-2N_{f}}.
\]

\noindent Parameters are $t_{C}=-2,\ t_{F}=0.5,\ \Lambda_{qcd}=592(\mathrm{MeV}),\ N_{c}=3,\ N_{f}=3$. (Since the value of $\Lambda_{qcd}$ is not important here, $\Lambda_{qcd}$ is used as a scale factor.) We assume that strange quark plays a role only in the running coupling and the running coupling has no a chemical potential.

The CJT effective potential (\ref{eq:cjt3}) at QCD is obtained by multiplying the number of colors $N_{c}=3$ and flavors $N_{f}=2$. The number of colors and flavors result from the trace in (\ref{eq:cjt1}) and (\ref{eq:cjt2}). To find a critical point for the chiral phase transition, we consider the difference between the Nambu-Goldstone phase ($B_{n}(\bm{p})\neq 0$) and the Wigner phase ($B_{n}(\bm{p})=0$). Thus, a critical point is determined by calculating,

\begin{align}
V(B&\neq 0)-V(B=0)\nonumber\\[0.21cm]
\vspace{1em}
=&-\displaystyle \frac{N_{c}N_{f}T}{\pi^{2}}\sum_{n}\int_{0}^{\infty}dy\ y^{2}\Big(\log\Big[\frac{-C_{n}^{2}(y)p_{0}^{2}+A_{n}^{2}(y)y^{2}+B_{n}^{2}(y)}{-C_{n}^{W2}(y)p_{0}^{2}+A_{n}^{W2}(y)y^{2}}\Big]\nonumber\\[0.21cm]
\vspace{1em}
&+p_{0}^{2}\displaystyle \Big(\frac{C_{n}(y)}{C_{n}^{2}(y)p_{0}^{2}-A_{n}^{2}(y)y^{2}-B_{n}^{2}(y)}-\frac{C_{n}^{W}(y)}{C_{n}^{W2}(y)p_{0}^{2}-A_{n}^{W2}(y)y^{2}}\Big)\nonumber\\[0.21cm]
\vspace{1em}
&-y^{2}\displaystyle \Big(\frac{A_{n}(y)}{C_{n}^{2}(y)p_{0}^{2}-A_{n}^{2}(y)y^{2}-B_{n}^{2}(y)}-\frac{A_{n}^{W}(y)}{C_{n}^{W2}(y)p_{0}^{2}-A_{n}^{W2}(y)y^{2}}\Big)\Big),
\end{align}

\noindent where $C_{n}^{W}$ and $A_{n}^{W}$ are solutions at the Wigner phase. If $V(B\neq 0)-V(B=0)\geq 0$, the chiral symmetry is restored \cite{rf:2}. Since there is a tricritical point at non-zero chemical potential, we especially investigated the effects on a tricritical point.

We used the iterative method to solve the SDE. For example, (\ref{eq:gene3}) on the iterative calculation is formally,

\begin{align}
B_{new}(x)=-T\displaystyle \frac{3e^{2}}{8\pi^{2}x}\sum_{m}\int_{0}^{\infty}dy&\displaystyle \frac{yB_{old}(y)}{C_{old}^{2}(y)q_{0}^{2}-A_{old}^{2}(y)y^{2}-B_{old}^{2}(y)}\log\frac{(p_{0}-q_{0})^{2}-(x+y)^{2}}{(p_{0}-q_{0})^{2}-(x-y)^{2}}
\end{align}

\vspace{1em}
\noindent It repeats until a value is converged. Then, we impose following restrictions:

\begin{description}

\item[(i):]\ repeat until a difference between $B_{new}$ and $B_{old}$ becomes $10^{-4}$ (MeV) order, and a difference between $C_{new}\ (A_{new})$ and $C_{old}\ (A_{old})$ become $10^{-7}$ order,

\vspace{1em}
\item[(ii):]\ on decision of a tricritical point, we consider $10^{-1}$(MeV) order for a temperature and a chemical potential,

\vspace{1em}
\item[(iii):\ ]on decision of a tricritical point, a point where $B_{n}(x)$ has a value of $10^{0}\sim 10^{-1}$ (MeV) order before $B_{n}(x)$ becomes zero numerically is a tricritical point,

\vspace{1em}
\item[(iv):]\ when $B_{n}(x)$ has a value of $10^{-2}$ (MeV) order, we regard $B_{n}(x)$ as zero numerically.

\end{description}

\noindent A tricritical point fluctuates somewhat by a numerical setup and precision. We note that a tricritical point obtained by our numerical calculation are not highly precise, because our main purpose is to study the effect of an imaginary part. For example, if we take $10^{-2}$ (MeV) for the order of $B_{n}(x)$ in (iii), the tricritical point moves to ($142,35$)MeV in the case (A1).

\vspace{1em}
\vspace{1em}
\subsection{Effect of the imaginary part}\label{sec:num1}

Since we want to know the effect of the imaginary part of the SDE, we try two cases: (A1) including the imaginary part, (B1) no including the imaginary part. When performing the numerical calculation, we use the real part and the imaginary part of the SDE in the case (A1), and use only the real part of the SDE in the case (B1) (fixing ${\rm Im} C_{n}(x), {\rm Im} A_{n}(x),$ and ${\rm Im} B_{n}(x)$ to zero). Then, inserting a solution in (A1) and (B1) into the CJT effective potential, we can verify the effect of an imaginary part for a tricritical point.

\begin{figure}[t]

\begin{tabular}{cc}

\begin{minipage}{0.48\hsize}

\begin{center}

\includegraphics[width=65mm]{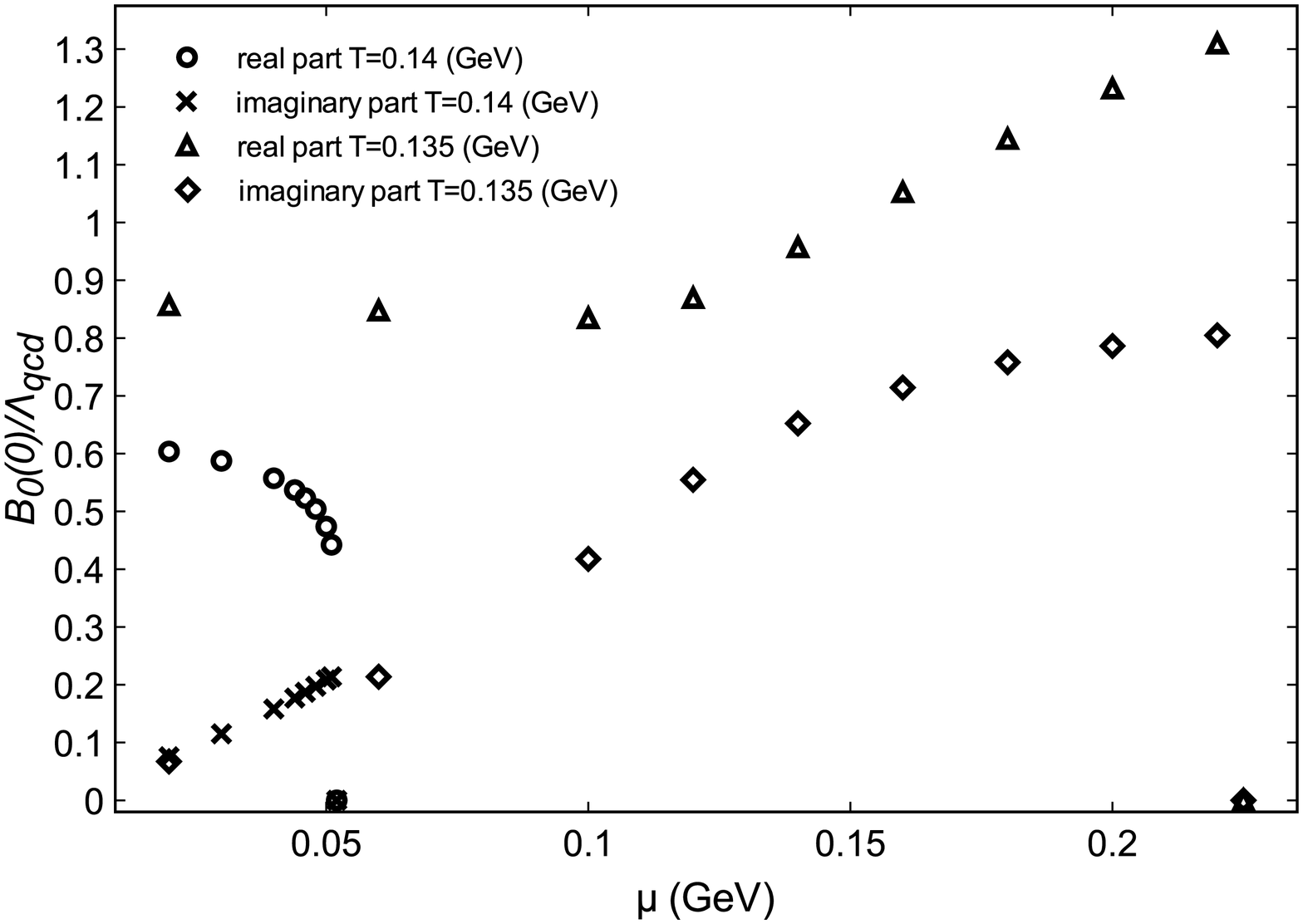}

\caption{The chemical potential dependence of $B_{0}(0)$ in (A1). $T=0.135,0.14$ (GeV).}

\label{fig:chemi}

\end{center}

\end{minipage}

\hspace{0.1cm}

\begin{minipage}{0.48\hsize}

\begin{center}

\includegraphics[width=65mm]{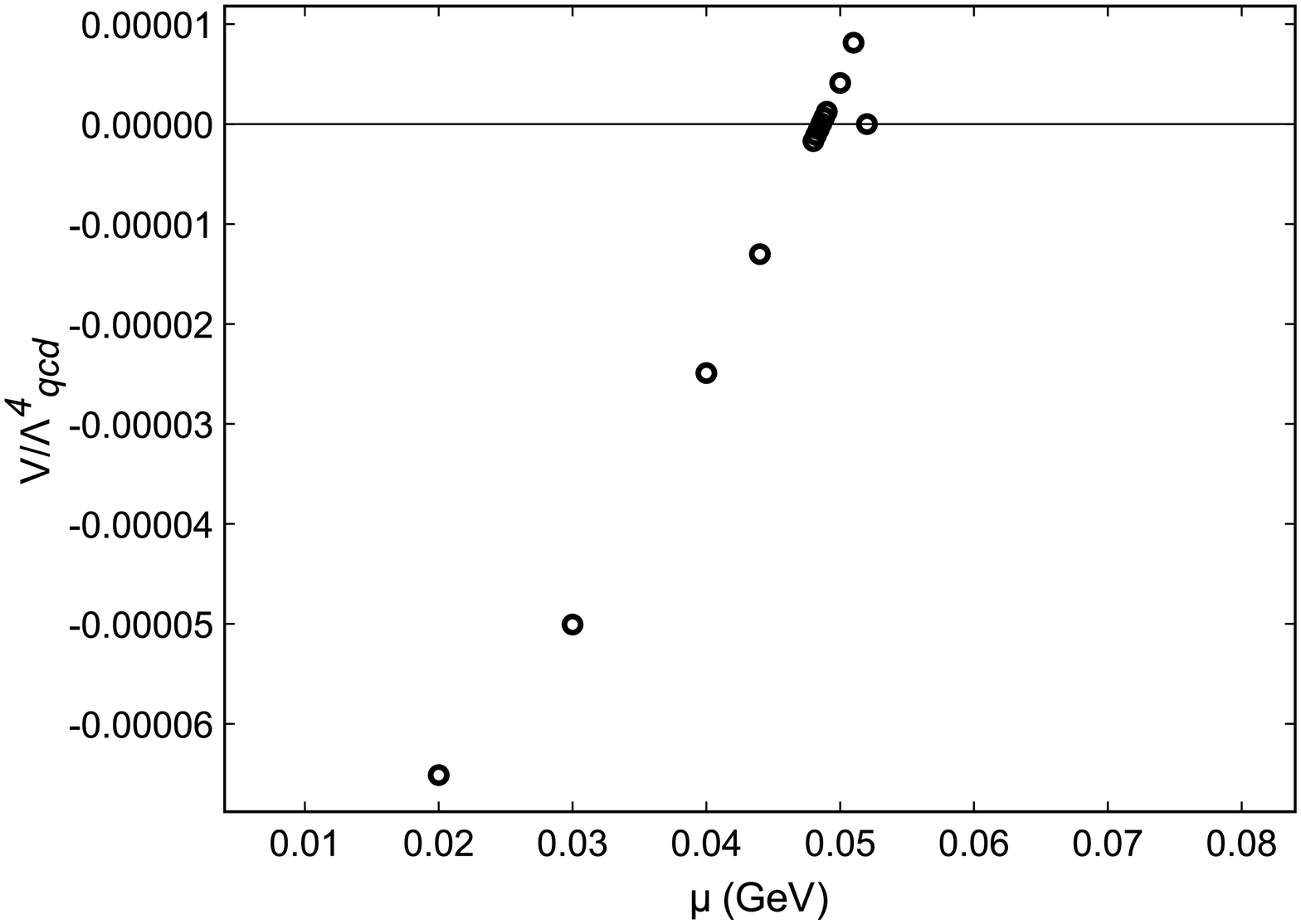}

\caption{The chemical potential dependence of $V(B\neq 0)-V(B=0)$ in (A1). $T=0.14$ (GeV). }

\label{fig:chemiV}

\end{center}

\end{minipage}

\end{tabular}

\end{figure}

\begin{figure}[t]

\begin{tabular}{cc}

\begin{minipage}{0.48\hsize}

\begin{center}

\includegraphics[width=65mm]{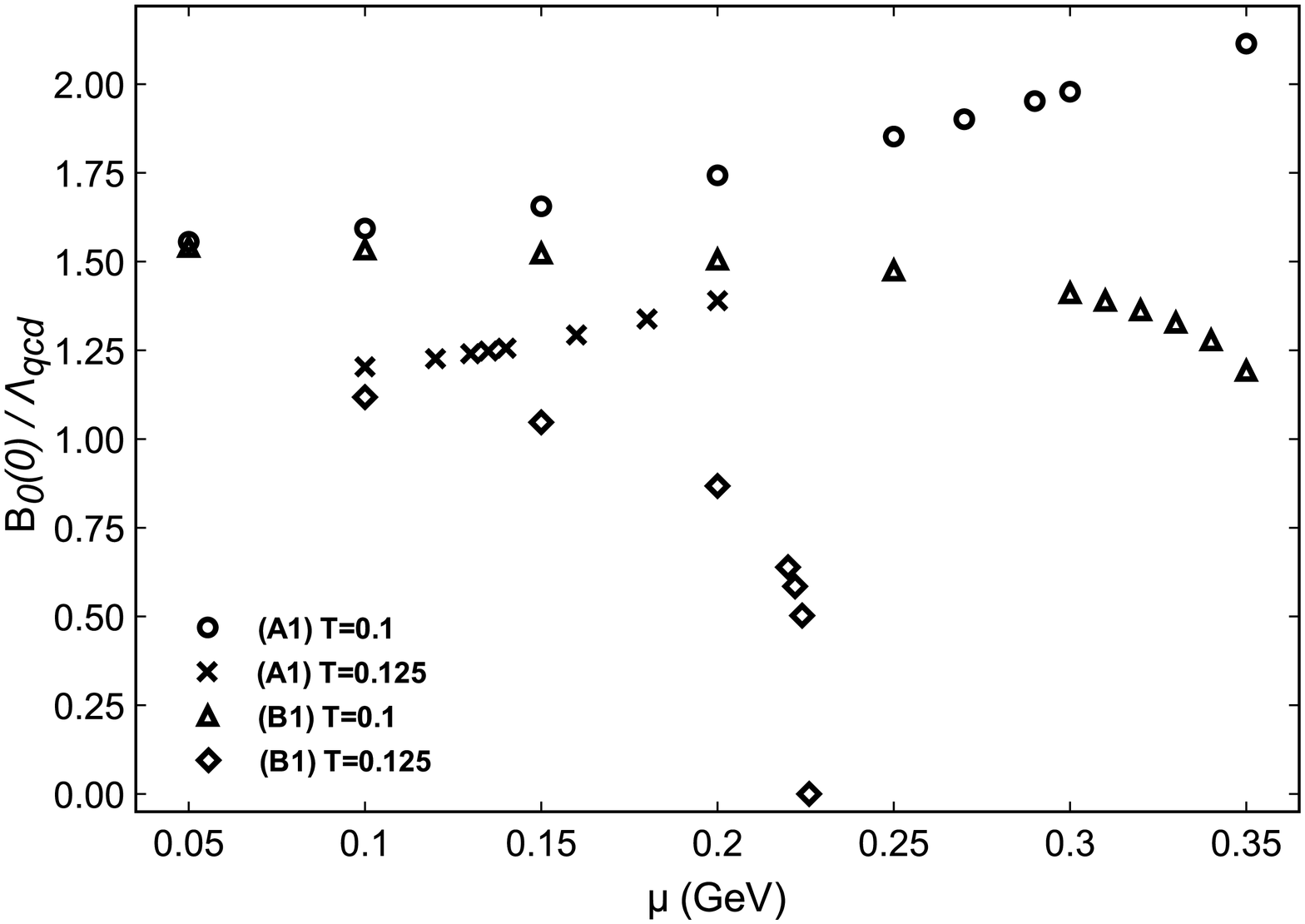}

\caption{The chemical potential dependence of ${\rm Re} B_{0}(0)$ in (A1) and (B1). To facilitate visualization, we depicted the results of $V(B\neq 0)-V(B=0)\leq 0$.}

\label{fig:gene}

\end{center}

\end{minipage}

\hspace{0.1cm}

\begin{minipage}{0.48\hsize}

\begin{center}

\includegraphics[width=65mm]{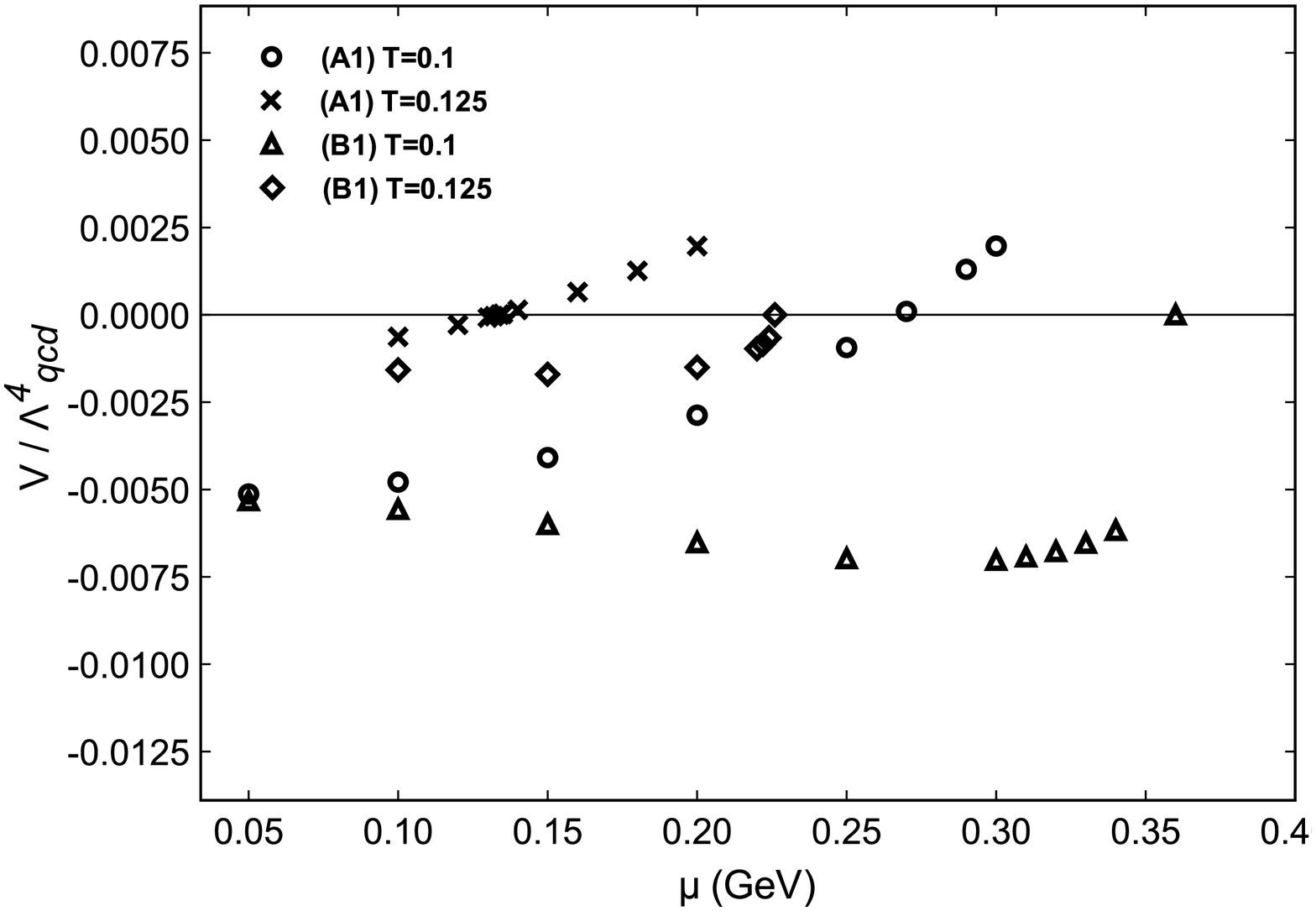}

\caption{The chemical potential dependence of $V(B\neq 0)-V(B=0)$ in (A1) and (B1).}

\label{fig:Vgene}

\end{center}

\end{minipage}

\end{tabular}

\end{figure}

The results for cases (A1) and (B1) are shown in Figs.\ \ref{fig:gene} and \ref{fig:Vgene}. The phase transition by a chemical potential is of first order, because $B_{n}(x)$ vanishes discontinuously (see Figs.\ \ref{fig:chemi} and \ref{fig:chemiV}). The tricritical point is $(143,28)$MeV in the case (A1), $(128,209)$MeV in the case (B1)  (see Fig.\ \ref{fig:tricri}).

\begin{figure}[t]

\begin{tabular}{cc}

\begin{minipage}{0.48\hsize}

\begin{center}

\includegraphics[width=65mm]{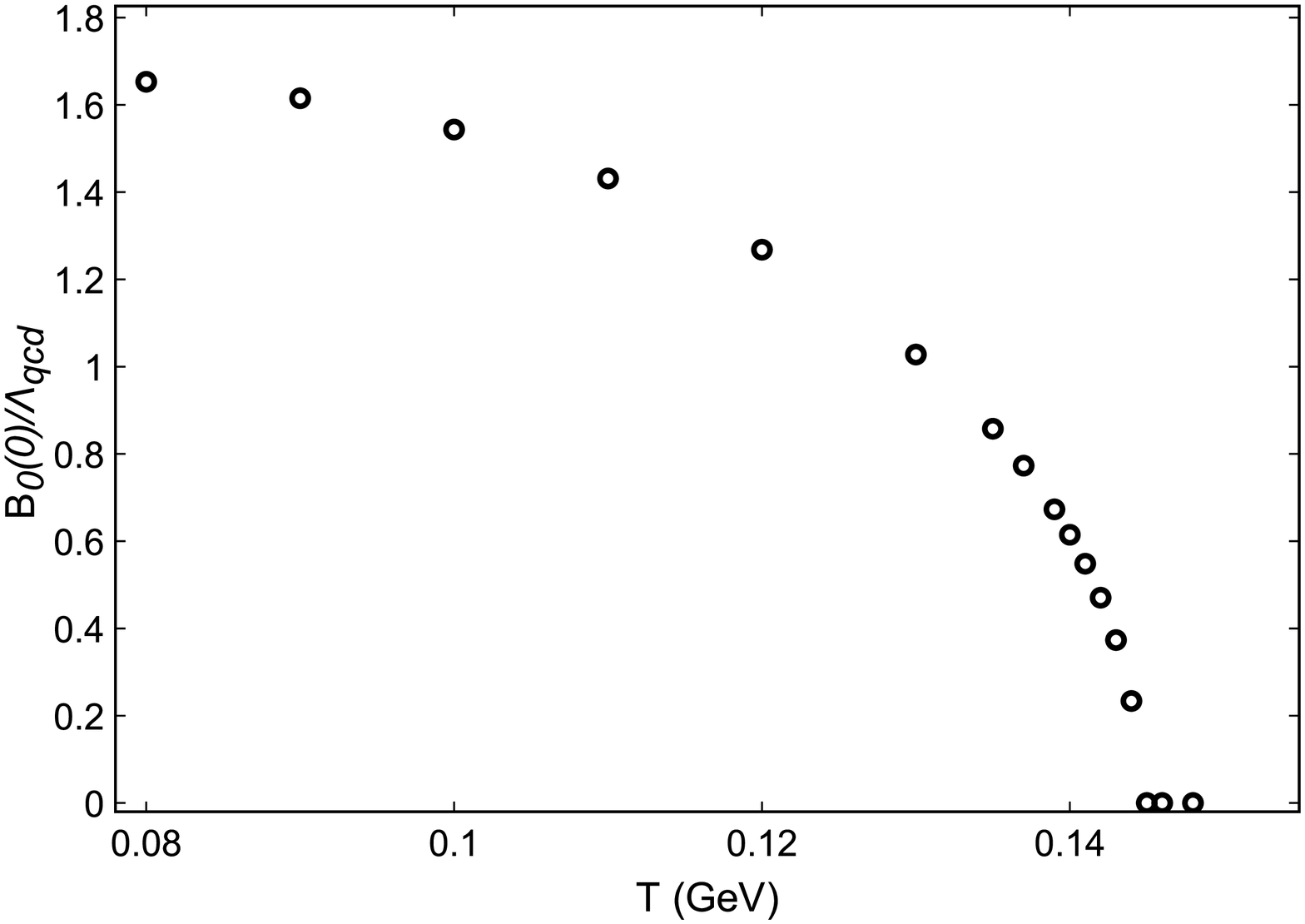}

\caption{The temperature dependence of $B_{0}(0)$ at $\mu=0$ in (A1).}

\label{fig:BT}

\end{center}

\end{minipage}

\hspace{0.1cm}

\begin{minipage}{0.48\hsize}

\begin{center}

\includegraphics[width=65mm]{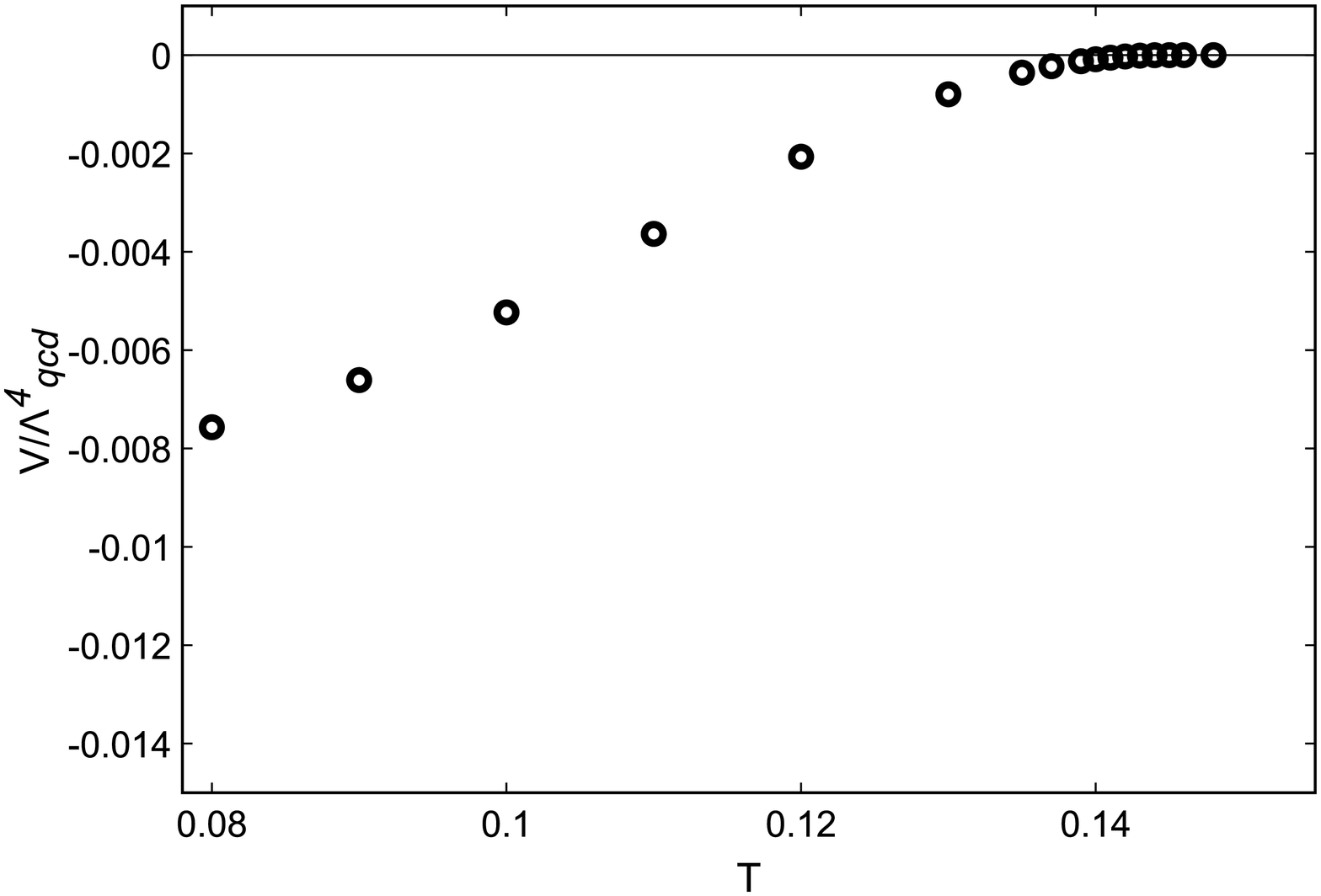}

\caption{The temperature dependence of $V(B\neq 0)-V(B=0)$ at $\mu=0$ in (A1).}

\label{fig:VT}

\end{center}

\end{minipage}

\end{tabular}

\end{figure}

The chemical potential dependence for the effective potential in the case (B1) behaves in the same way as the temperature dependence. There is a region where the effective potential $V(B_{n}(x)\neq 0)-V(B_{n}(x)=0)$ does not become positive by increasing a chemical potential (Fig.\ \ref{fig:VT}). Although $B_{n}(x)$ decreases smoothly by increasing a chemical potential, $B_{n}(x)$ vanishes discontinuously (Fig.\ \ref{fig:BT}). (When non-zero physical quark masses are used, the chiral transition at zero chemical potential becomes a crossover \cite{rf:gavai}.) Even so, the phase transition is of first order, and a tricritical point exists. In constrast, although the chemical potential dependence of $B_{n}(x)$ in the case (A1) behaves like the temperature dependence at around the tricritical point, $V(B_{n}(x)\neq 0)-V(B_{n}(x)=0)$ has a positive value (see Figs.\ \ref{fig:chemi} and \ref{fig:chemiV}).

As a result, the imaginary part affects the phase transition, and enhances the effect of the first order transition. In addition, the critical chemical potential is displaced by the imaginary part (about $200$ (MeV)).

\vspace{1em}
\subsection{Effect of the wave function renormalization constant}\label{sec:num2}

In place of (\ref{eq:pro1}), we use the exact fermion green function,

\begin{equation}
G_{\beta}(p)=\displaystyle \frac{-1}{\gamma_{0}p_{0}+\gamma_{i}p^{i}-B_{n}(\bm{p})}.\label{eq:pro2}
\end{equation}

\noindent This is the approximate form by $C_{n}(\bm{p})=A_{n}(\bm{p})=1$, and this is used from the analogy of zero temperature. (Adopting the Landau gauge, $C(p)$ and $A(p)$ are $1$ at zero temperature and chemical potential. However, $C_{n}(\bm{p})$ and $A_{n}(\bm{p})$ are not $1$ at non-zero temperature and chemical potential \cite{rf:ss}.) The effect of the wave function renormalization constant $C_{n}(\mbox{\boldmath $p$})$ and $A_{n}(\mbox{\boldmath $p$})$ is verified by comparing (\ref{eq:pro1}) to (\ref{eq:pro2}). Thus, we calculate two cases:

\begin{description}

\item[(A2)]:\ the SDE includes the imaginary part and has no wave function renormalization constant,

\vspace{1em}
\item[(B2)]:\ the SDE does not include the imaginary part and has no wave function renormalization constant.

\end{description}

\begin{figure}[t]

\begin{tabular}{cc}

\begin{minipage}{0.48\hsize}

\begin{center}

\includegraphics[width=65mm]{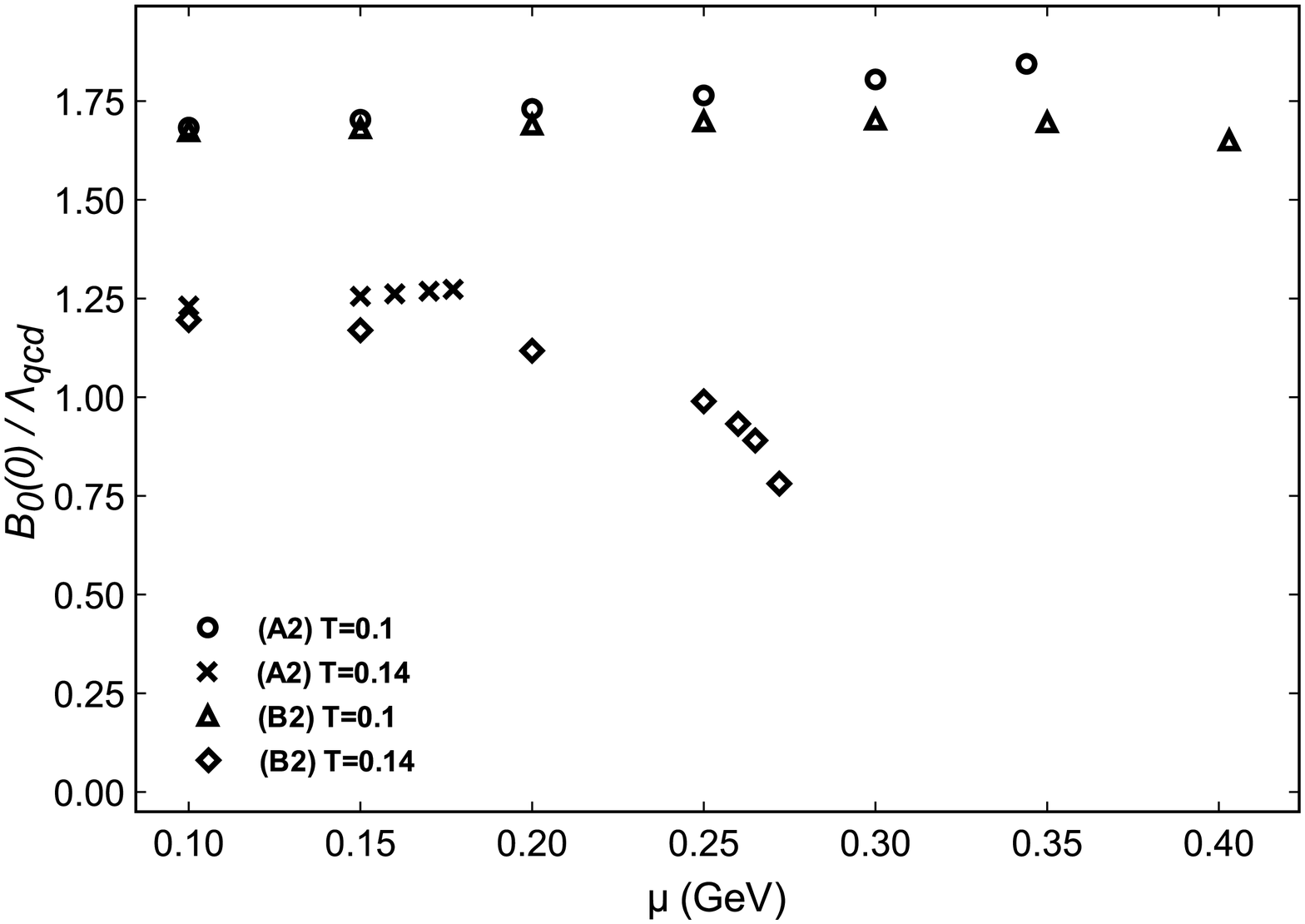}

\caption{The chemical potential dependence of ${\rm Re} B_{0}(0)$ in (A2) and (B2).}

\label{fig:onlyB}

\end{center}

\end{minipage}

\hspace{0.1cm}

\begin{minipage}{0.48\hsize}

\begin{center}

\includegraphics[width=65mm]{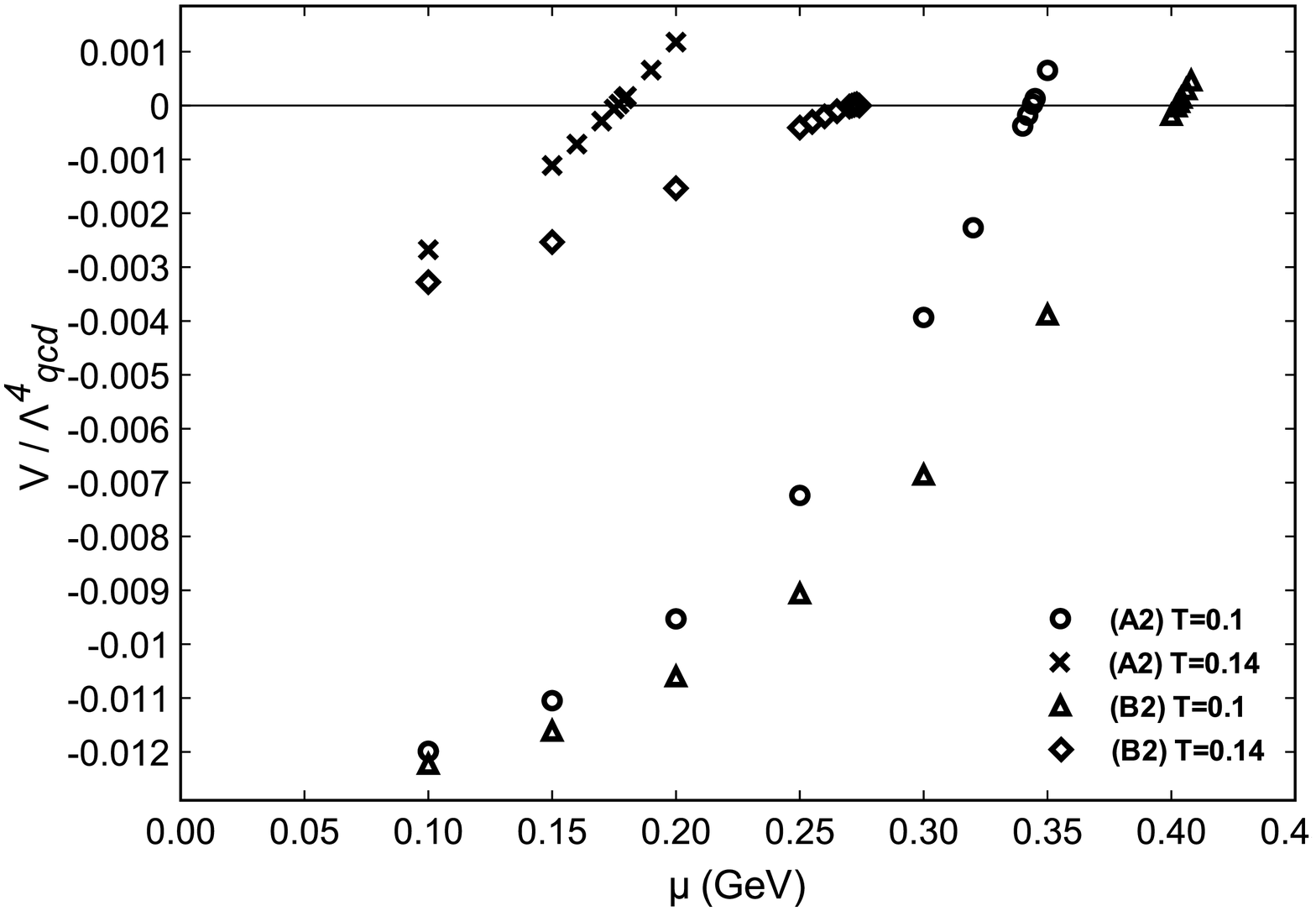}

\caption{The chemical potential dependence of $V(B\neq 0)-V(B=0)$ in (A2) and (B2).}

\label{fig:VonlyB}

\end{center}

\end{minipage}

\end{tabular}

\end{figure}

\begin{figure}[t]

\begin{tabular}{cc}

\begin{minipage}{0.5\hsize}

\begin{center}

\includegraphics[width=65mm]{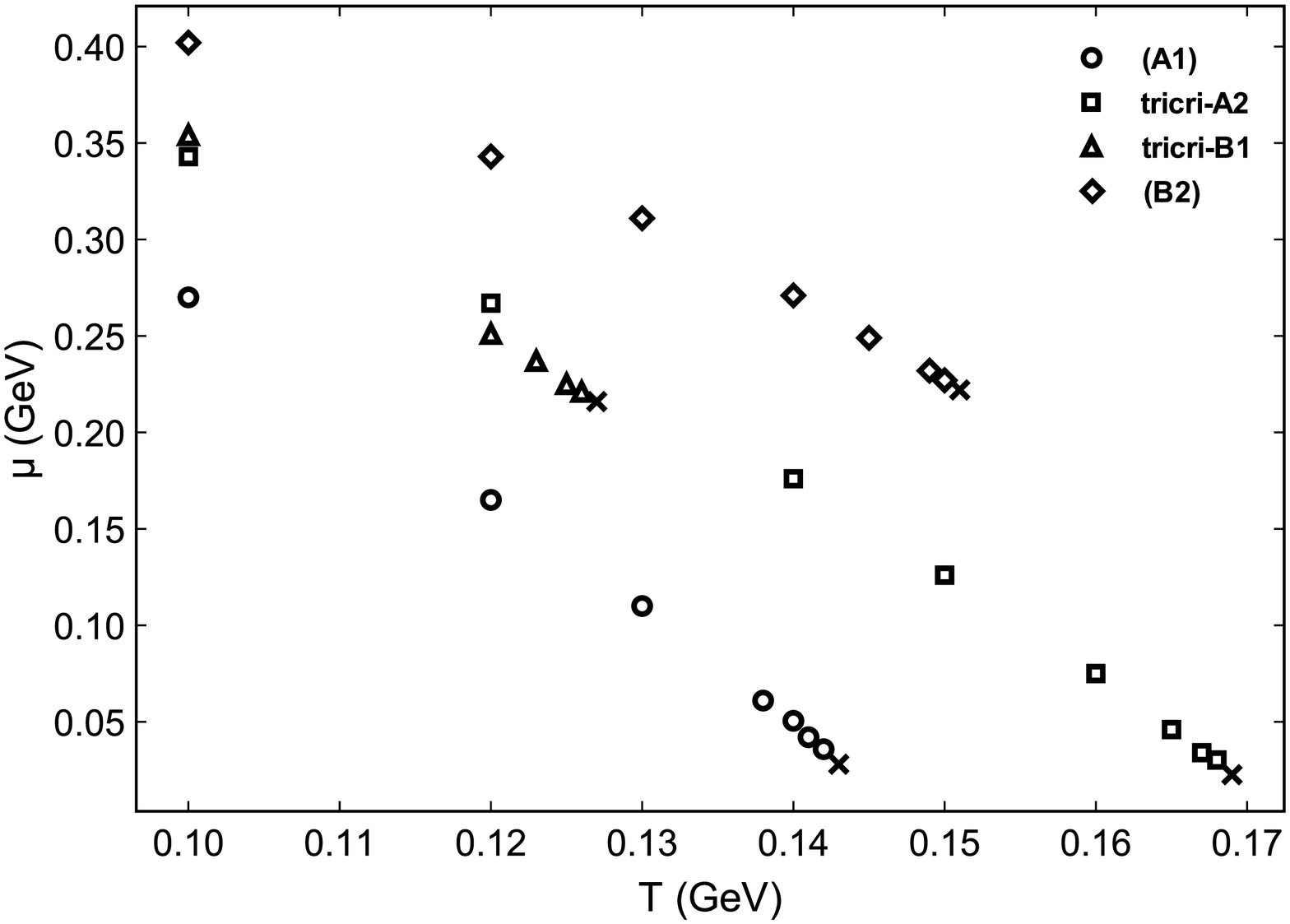}

\caption{Tricritical points in cases (A1), (A2), (B1), and (B2). Cross marks indicate each tricritical points. Others are the phase transition points of first order.}

\label{fig:tricri}

\end{center}

\end{minipage}

\hspace{0.1cm}

\begin{minipage}{0.5\hsize}

\begin{center}

\includegraphics[width=65mm]{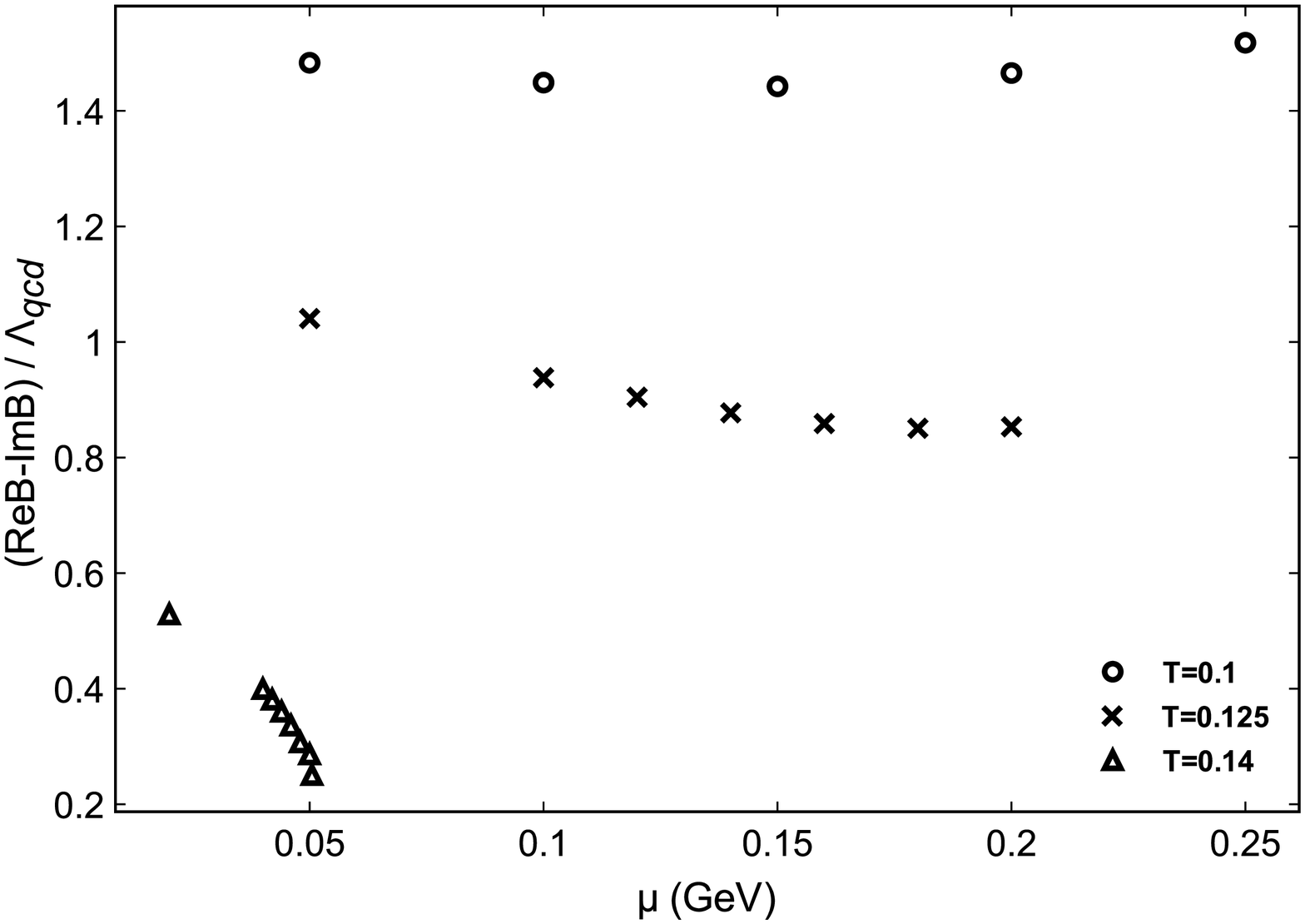}

\caption{The difference of ${\rm Re} B_{0}(0)$ and ${\rm Im} B_{0}(0)$. Critical points are ($100,268$), ($125,133$) and ($140,48$)MeV.\newline}

\label{fig:Br-Bi}

\end{center}

\end{minipage}

\end{tabular}

\end{figure}

The results by using (\ref{eq:pro2}) are shown in Figs.\ \ref{fig:onlyB} and \ref{fig:VonlyB}. The chemical potential dependence in the cases (A2) and (B2) behaves in the same way as cases (A1) and (B1). Then, as shown in Fig.\ \ref{fig:tricri}, the chemical potential at the tricritical point in the case (B2) is much larger than that of (A2): ($ T,\mu$)=($170,15$)MeV in the case (A2), ($151,222$)MeV in the case (B2). In contrast to this, the temperature of the tricritical point in the case (B2) is slightly smaller than that of the case (A2). As a result, in the case (A1) ((B1)), $C_{n}(x)$ and $A_{n}(x)$ decrease the temperature of the tricritical point, and increase (decrease) the chemical potential of the tricritical point.

\begin{figure}[t]

\begin{center}

\includegraphics[width=65mm]{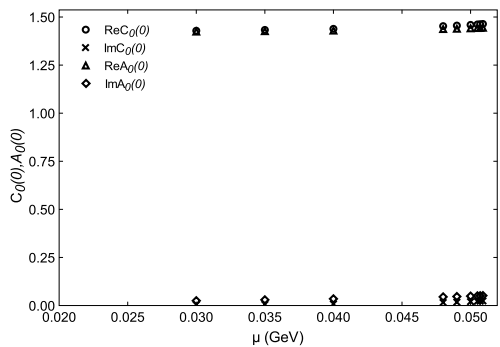}

\caption{The chemical potential dependence of $C_{0}(0)$ and $A_{0}(0).\ T=0.14$ (GeV).}

\label{fig:chemiCA}

\end{center}

\end{figure}

The clear effect of the wave function renormalization constant for the tricritical point is to lower the critical temperature. The difference is about $20$ (MeV). On the other hand, since the effect of the wave function renormalization constant for a critical chemical potential is different in (A) and (B), the effect of their imaginary part for a critical chemical potential should be strong. However, Fig.\ \ref{fig:chemiCA} shows that ${\rm Im} C_{n}(x)$ and ${\rm Im} A_{n}(x)$ is very small at around the tricritical point. In constrast, ${\rm Im} B_{n}(x)$ has a value of the same order to ${\rm Re} B_{n}(x)$ at around the tricritical point, and has a strong effect (comparing the results of (A2) and (B2)). Thus, we expect that main cause of this difference arises from $B_{n}(x)$.

\vspace{1em}

\vspace{1em}

\vspace{1em}

\vspace{1em}

\vspace{1em}

\vspace{1em}
\vspace{1em}
\vspace{1em}
\section{summary and disucussion}\label{sec:sum}

In this paper, we calculated the improved ladder approximation SDE at non-zero temperature and chemical potential, and we verified the effect of the imaginary part and the wave function renormalization constant in the SDE. Considering the imaginaty part and the wave function renormalization constant, all effects in the ladder approximation are included. Thus, we calculate four cases:

\begin{description}

\item[(A1)]:\ the SDE includes the imaginary part and has the wave function renormalization constant,

\vspace{1em}
\item[(A2)]:\ the SDE includes the imaginary part and has no wave function renormalization constant,

\vspace{1em}
\item[(B1)]:\ the SDE does not include the imaginary part and has the wave function renormalization constant,

\vspace{1em}
\item[(B2)]:\ the SDE does not include the imaginary part and has no wave function renormalization constant.

\end{description}

\noindent In particular, we took notice of a tricritical point. The tricritical point is ($ T,\mu$)=$(143,28)$MeV in the case (A1), ($170,15$)MeV in the case (A2), ($128,209$)MeV in the case (B1), and ($151,222$)MeV in the case (B2).

The tricritical point in \cite{rf:3} is ($142,82$)MeV in the case (A1) and ($210,43$)MeV in the case (A2). The difference should result from using a different running coupling constant and a procedure on a numerical calculation. (From the result in the case (A1), although it seems that the temperature at a tricritical point does not dependent on a choice of a running coupling constant, it should be a coincidence. Because, (A2) disagrees with our result.) The property, that the temperature decreases and the chemical potential increases to use (\ref{eq:pro1}) in place of (\ref{eq:pro2}), corresponds with our result. Thus, $C_{n}(x)$ and $A_{n}(x)$ decrease the temperature, and increase (decrease) the chemical potential at the tricritical point in the case (A) ((B)).

On the other hand, although a tricritical point is moved by $C_{n}(x)$ and $A_{n}(x)$, it is small (about $10$ (MeV)). Furthermore, $C_{n}(x)$ and $A_{n}(x)$ should not affect the properties of the phase transition strongly. For example, they do not alter the order of phase transition. Thus, when one studies the chiral phase transition, it is sufficient to be able to use $C_{n}(x)=A_{n}(x)=1$. (In addition, their imaginary parts are very small.)

The chemical potential dependence of $B_{n}(x)$ and the effective potential in the case (B) behaves like the temperature dependence. When there are not imaginary parts, the effect of temperature, that produces the second order phase transition, should be stronger. In addition, the chemical potential of the tricritical point in the case (B) is larger than the case (A) (about $200$ (MeV)). As a result, we expect that the imaginary part affects the phase transition and enhances the effect of the first order transition.

For both cases (A) and (B), the imaginary part moves to the chemical potential largely (above $100$ (MeV)) at the tricritical point. These results for both cases are the same as \cite{rf:4}. In particular, ${\rm Im} B_{n}(x)$ contributes to a tricritical point strongly. Thus, when one studies a tricritical point, one should not ignore an imaginary part in the SDE. On the other hand, the contribution of ${\rm Im} B_{n}(x)$ is small at a point away from the tricritical point. For example, the critical point is ($100,268$)(MeV) in the case (A1), and is ($100,355$)(MeV) in the case (B1). This is understood from the fact that the difference between ${\rm Re} B_{n}(x)$ and ${\rm Im} B_{n}(x)$ gets larger at low temperature (see Fig.\ \ref{fig:Br-Bi}). Thus, we should be able to ignore the imaginary part at a point away from a tricritical point, because the imaginary part is very small. However, as mentioned above, the chemical potential dependence of the effective potential becomes as the temperature dependence by ignoring the imaginary part. Due to this, although one can ignore the imaginary part in detemining a critical point approximately, one should not ignore the imaginary part in studying a property of the phase transition.

Althogh we considered only the chiral phase transition, there is the deconfinement phase transition in hot and dense QCD. These relations are less well understood. For example, the critical temperature is different in \cite{rf:lat1}. On the other hand, in \cite{rf:lat2}, the critical temperature is the same. In addition, the analysis at non-zero chemical potential is more uncertain. We show the result in appendix\ B. The behavior of $B_{n}(x)$ in the chiral limit is identical with \cite{rf:deco}. The critical temperature for the chiral phase transition should coincide with the deconfinment transition within the error. On the other hand, the critical chemical potential has a gap even in the chiral limit. It is necessary to study the relation between the chiral and the deconfinment phase transition at non-zeto chemical potential.

\vspace{1em}
\vspace{1em}
\vspace{1em}
\appendix

\section{$I,\ H$, real parts, and imaginary parts in the SDE }\label{app1}

We show explicit expressions of $I$ and $H$ in (\ref{eq:gene1})--(\ref{eq:gene3}).

\vspace{1em}
\[
a_{+}=(p_{0}-q_{0})^{2}-(x+y)^{2}\ ,\ a_{-}=(p_{0}-q_{0})^{2}-(x-y)^{2}.
\]

\begin{itemize}

\item$C_{n}(x)$

\vspace{1em}
\[
I_{1}=2q_{0}\log\frac{a_{+}}{a_{-}}\ ,\ I_{2}=q_{0}\Big(-2(p_{0}-q_{0})^{2}\Big(\frac{1}{a_{+}}-\frac{1}{a_{-}}\Big)-\log\frac{a_{+}}{a_{-}}\Big),
\]

\vspace{1em}
\[
I_{3}=-(p_{0}-q_{0})\Big[\log\frac{a_{+}}{a_{-}}-\Big(-(p_{0}-q_{0})^{2}+x^{2}-y^{2})\Big(\frac{1}{a_{+}}-\frac{1}{a_{-}}\Big)\Big].
\]

\vspace{1em}
\item$A_{n}(x)$

\vspace{1em}
\[
H_{1}=(p_{0}-q_{0})q_{0}\Big[(x^{2}-y^{2}+(p_{0}-q_{0})^{2})\Big(\frac{1}{a_{+}}-\frac{1}{a_{-}}\Big)+\log\frac{a_{+}}{a_{-}}\Big],
\]

\vspace{1em}
\[
H_{2}=-4xy+(x^{2}+y^{2}-(p_{0}-q_{0})^{2})\log\frac{a_{+}}{a_{-}},
\]

\begin{align*}
H_{3}=\displaystyle \Big(x^{2}+y^{2}-&\displaystyle \frac{x^{2}+y^{2}-(p_{0}-q_{0})^{2}}{2}\Big)\log\frac{a_{+}}{a_{-}}\\[0.28cm]
\vspace{1em}
&-\displaystyle \Big(\frac{(x^{2}-y^{2})^{2}-(p_{0}-q_{0})^{4}}{2}\Big)\Big(\frac{1}{a_{+}}-\frac{1}{a_{-}}\Big).
\end{align*}

\end{itemize}

\vspace{1em}
The real parts and imaginary parts in (\ref{eq:gene1})--(\ref{eq:gene3}) are

\vspace{1em}
\begin{align*}
&{\rm Re} C_{n}(x)=1+\displaystyle \frac{e^{2}T}{8\pi^{3}x}\frac{1}{(\omega_{n}^{2}+\mu^{2})}\sum_{m}\int_{0}^{\infty}dyy[(I_{1}^{\prime}+I_{2}^{\prime})(u_{1}v_{1}-u_{2}v_{2})+I_{3}^{\prime}(v_{1}{\rm Re} A_{m}-v_{2}{\rm Im} A_{m})]\frac{1}{R_{1}^{2}+R_{2}^{2}},\\[0.21cm]
\vspace{1em}
&{\rm Im} C_{n}(x)=\displaystyle \frac{e^{2}T}{8\pi^{3}x}\frac{1}{(\omega_{n}^{2}+\mu^{2})}\sum_{m}\int_{0}^{\infty}dyy[(I_{1}^{\prime}+I_{2}^{\prime})(u_{2}v_{1}+u_{1}v_{2})+I_{3}^{\prime}(v_{2}{\rm Re} A_{m}+v_{1}{\rm Im} A_{m})]\frac{1}{R_{1}^{2}+R_{2}^{2}},\\[0.21cm]
\vspace{1em}
&{\rm Re} A_{n}(x)=1+\displaystyle \frac{e^{2}T}{8\pi^{3}x^{3}}\sum_{m}\int_{0}^{\infty}dyy[-H_{1}^{\prime}(u_{1}R_{1}+u_{2}R_{2})+(H_{2}-H_{3})(R_{1}{\rm Re} A+R_{2}{\rm Im} A)]\frac{1}{R_{1}^{2}+R_{2}^{2}},\\[0.21cm]
\vspace{1em}
&{\rm Im} A_{n}(x)=\displaystyle \frac{e^{2}T}{8\pi^{3}x^{3}}\sum_{m}\int_{0}^{\infty}dyy[-H_{1}^{\prime}(u_{2}R_{1}-u_{1}R_{2})+(H_{2}-H_{3})(R_{1}{\rm Im} A-R_{2}{\rm Re} A)]\frac{1}{R_{1}^{2}+R_{2}^{2}},\\[0.21cm]
\vspace{1em}
&{\rm Re} B_{n}(x)=\displaystyle \frac{3e^{2}T}{8\pi^{2}x}\sum_{m}\int_{0}^{\infty}dyy\frac{R_{1}{\rm Re} B_{m}+R_{2}{\rm Im} B_{m}}{R_{1}^{2}+R_{2}^{2}}\log\frac{a_{+}}{a_{-}},\\[0.21cm]
\vspace{1em}
&{\rm Im} B_{n}(x)=\displaystyle \frac{3e^{2}T}{8\pi^{2}x}\sum_{m}\int_{0}^{\infty}dyy\frac{R_{1}{\rm Im} B_{m}-R_{2}{\rm Re} B_{m}}{R_{1}^{2}+R_{2}^{2}}\log\frac{a_{+}}{a_{-}}
\end{align*}

\noindent where

\begin{align*}
R_{1}=(\omega_{m}^{2}-\mu^{2})({\rm Re} C_{m}^{2}(y)-{\rm Im} C_{m}^{2}(y))&+4\mu\omega_{m}{\rm Re} C_{m}(y){\rm Im} C_{m}(y)\\[0.21cm]
\vspace{1em}
&+y^{2}({\rm Re} A_{m}^{2}(y)-{\rm Im} A_{m}^{2}(y))+({\rm Re} B_{m}^{2}(y)-{\rm Im} B_{m}^{2}(y))
\end{align*}

\begin{align*}
R_{2}=2(\omega_{m}^{2}-\mu^{2}){\rm Re} C_{m}(y){\rm Im} C_{m}(y)&-2\mu\omega_{m}({\rm Re} C_{m}^{2}(y)-{\rm Im} C_{m}^{2}(y))\\[0.21cm]
\vspace{1em}
&+2y^{2}{\rm Re} A_{m}(y){\rm Im} A_{m}(y)+2{\rm Re} B_{m}(y){\rm Im} B_{m}(y)
\end{align*}

\begin{center}

$u_{1}=\omega_{m}{\rm Re} C_{m}(y)+\mu{\rm Im} C_{m}(y),\ u_{2}=\omega_{m}{\rm Im} C_{m}(y)-\mu{\rm Re} C_{m}(y)$

\vspace{1em}
$v_{1}=\omega_{n}R_{1}+\mu R_{2},\ v_{2}=\mu R_{1}-\omega_{n}R_{2}$

\vspace{1em}
$I_{1}^{\prime}=2\displaystyle \log\frac{a_{+}}{a_{-}},\ I_{2}^{\prime}=-2(p_{0}-q_{0})^{2}\Big(\frac{1}{a_{+}}-\frac{1}{a_{-}}\Big)-\log\frac{a_{+}}{a_{-}}$

\vspace{1em}
$I_{3}^{\prime}=-(\displaystyle \omega_{n}-\omega_{m})\Big[\log\frac{a_{+}}{a_{-}}-((\omega_{n}-\omega_{m})^{2}+x^{2}-y^{2})\Big(\frac{1}{a_{+}}-\frac{1}{a_{-}}\Big)\Big]$

\vspace{1em}
$H_{1}^{\prime}=-(\displaystyle \omega_{n}-\omega_{m})\Big[(x^{2}-y^{2}-(\omega_{n}-\omega_{m})^{2})\Big(\frac{1}{a_{+}}-\frac{1}{a_{-}}\Big)+\log\frac{a_{+}}{a_{-}}\Big]$

\end{center}

\vspace{1em}
\vspace{1em}
\section{Deconfinement phase transition}

We use the dual quark condensate as an order parameter for center symmetry. The dual quark condensate is defined by \cite{rf:dress}

\vspace{1em}
\begin{equation}
\displaystyle \Sigma_{n}=\int_{0}^{2\pi}\frac{d\phi}{2\pi}e^{-i\phi n}\langle\overline{\psi}\psi\rangle_{\phi}.
\end{equation}

\vspace{1em}
\noindent$\langle\overline{\psi}\psi\rangle_{\phi}$ is given by

\vspace{1em}
\begin{align*}
\displaystyle \langle\overline{\psi}\psi\rangle_{\phi}=\ &N_{c}T\displaystyle \sum_{n}\int\frac{d^{3}p}{(2\pi)^{3}}\mathrm{tr}G_{\beta}(\omega_{n}(\phi),\bm{p})\nonumber\\[0.21cm]
\vspace{1em}
=\displaystyle \ &\frac{2N_{c}T}{\pi^{2}}\sum_{n}\int_{0}^{\infty}dx\frac{x^{2}B_{n}(x)}{C_{n}^{2}(x)(\omega_{n}(\phi)-i\mu)+A_{n}^{2}(x)x^{2}+B_{n}^{2}(x)},
\end{align*}

\vspace{1em}
\noindent where $\omega_{n}(\phi)=2\pi T(n+\phi/2\pi).\ \phi/2\pi$ is caused by the boundary condition $\psi(\beta,\bm{x})=e^{i\phi}\psi(0,\bm{x}). \Sigma_{1}$, whch is called the dressed Polyakov loop, contains the Polyakov loop. Thus, $\Sigma_{+1}$ (or $\Sigma_{-1}$) is the order parametr for deconfinement.

\begin{figure}[t]

\begin{center}

\includegraphics[width=65mm]{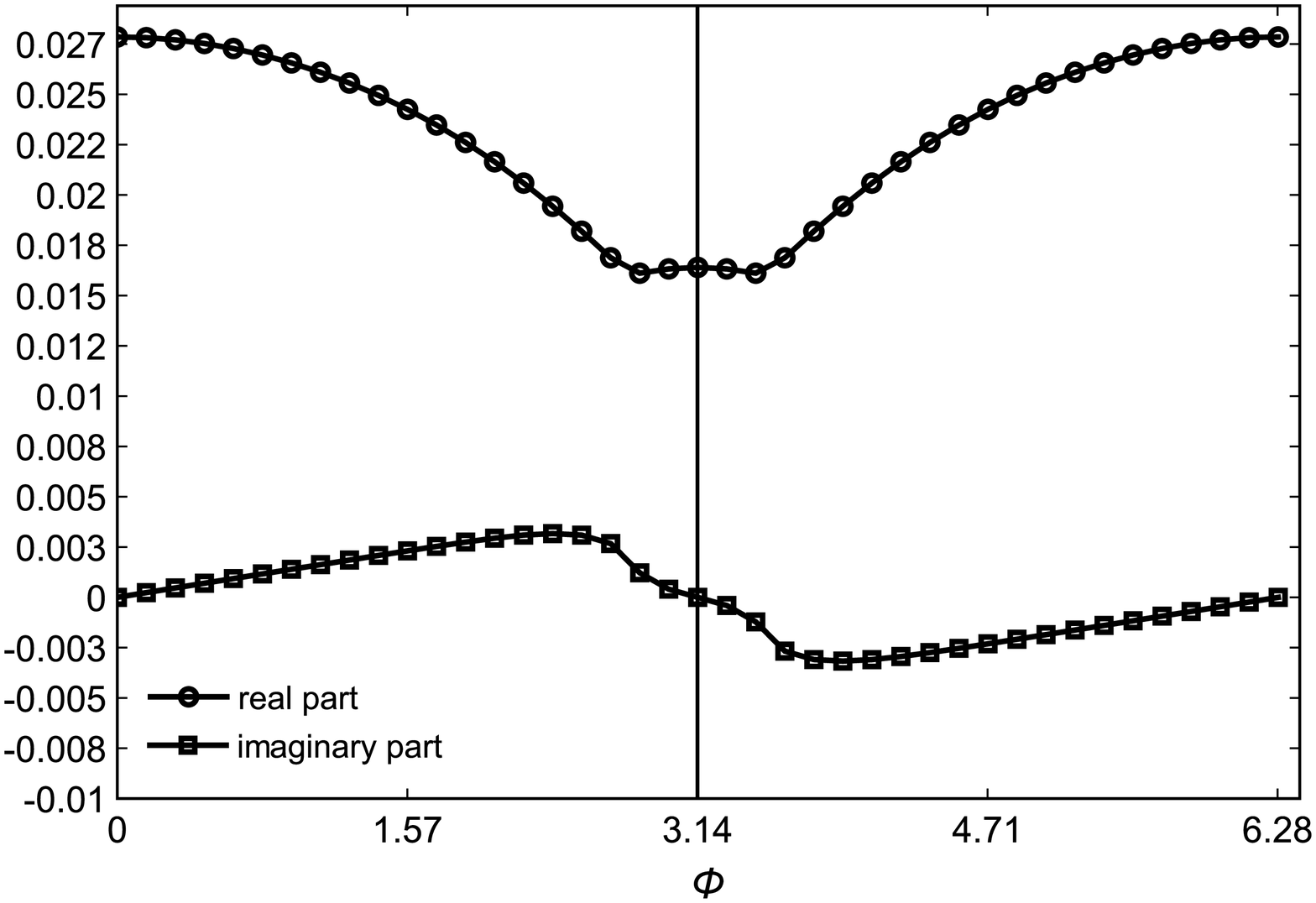}

\caption{The angular dependence of ${\rm Re}\langle\overline{\psi}\psi\rangle_{\phi}$ and ${\rm Im}\langle\overline{\psi}\psi\rangle_{\phi}.\ T=150$ (MeV), $\mu=120$( MeV).}

\label{fig:pol3}

\end{center}

\end{figure}

Although $\langle\overline{\psi}\psi\rangle_{\phi=\pi}$ has no imaginary part, $\langle\overline{\psi}\psi\rangle_{\phi\neq\pi}$ has a imaginary part at non-zero chemical potential. The real part is symmetric and the imaginary part is anti-symmetric (see Fig.\ \ref{fig:pol3}). For this reason, $\Sigma_{\pm 1}$ becomes

\vspace{1em}
\begin{equation}
\displaystyle \Sigma_{\pm 1}=\int_{0}^{2\pi}\frac{d\phi}{2\pi}({\rm Re}\langle\overline{\psi}\psi\rangle_{\phi}\cos\phi\pm{\rm Im}\langle\overline{\psi}\psi\rangle_{\phi}\sin\phi).
\end{equation}

\vspace{1em}
\noindent The second term vanishes at zero chemical potential.

To find a underlying property, we consider a simple approximation. Thus, we use the ladder approximation, $C_{n}(x)=A_{n}(x)=1$, and the chiral limit. Determing the critical temperature for deconfinement, we use \cite{rf:deco}

\vspace{1em}
\begin{equation}
\displaystyle \tau_{\pm 1}=\frac{1}{T^{2}}\frac{\partial\Sigma_{\pm 1}}{\partial T}.
\end{equation}

\begin{figure}[t]

\begin{tabular}{cc}

\begin{minipage}{0.5\hsize}

\begin{center}

\includegraphics[width=65mm]{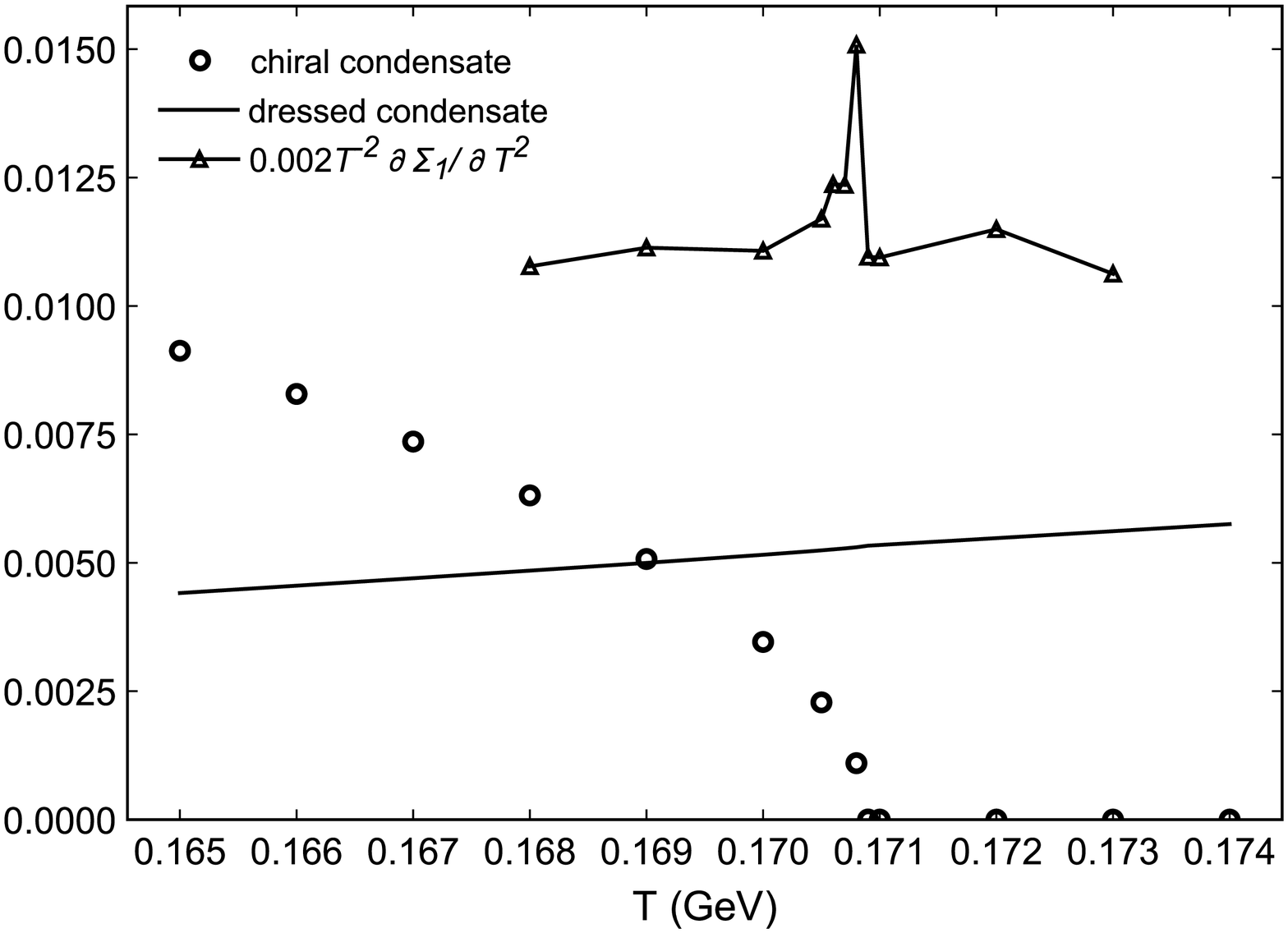}

\caption{The temperature dependence of the chiral condensate $\langle\overline{\psi}\psi\rangle_{\pi}$, the dressed Polyakov loop $\Sigma_{1}$, and $T^{-2}\partial\Sigma_{1}/\partial T.\ \langle\overline{\psi}\psi\rangle_{\pi}=0$ and the peak of $T^{-2}\partial\Sigma_{1}/\partial T$ is identical within $1$ (MeV).\newline}

\label{fig:pol1}

\end{center}

\end{minipage}

\hspace{0.1cm}

\begin{minipage}{0.5\hsize}

\begin{center}

\includegraphics[width=65mm]{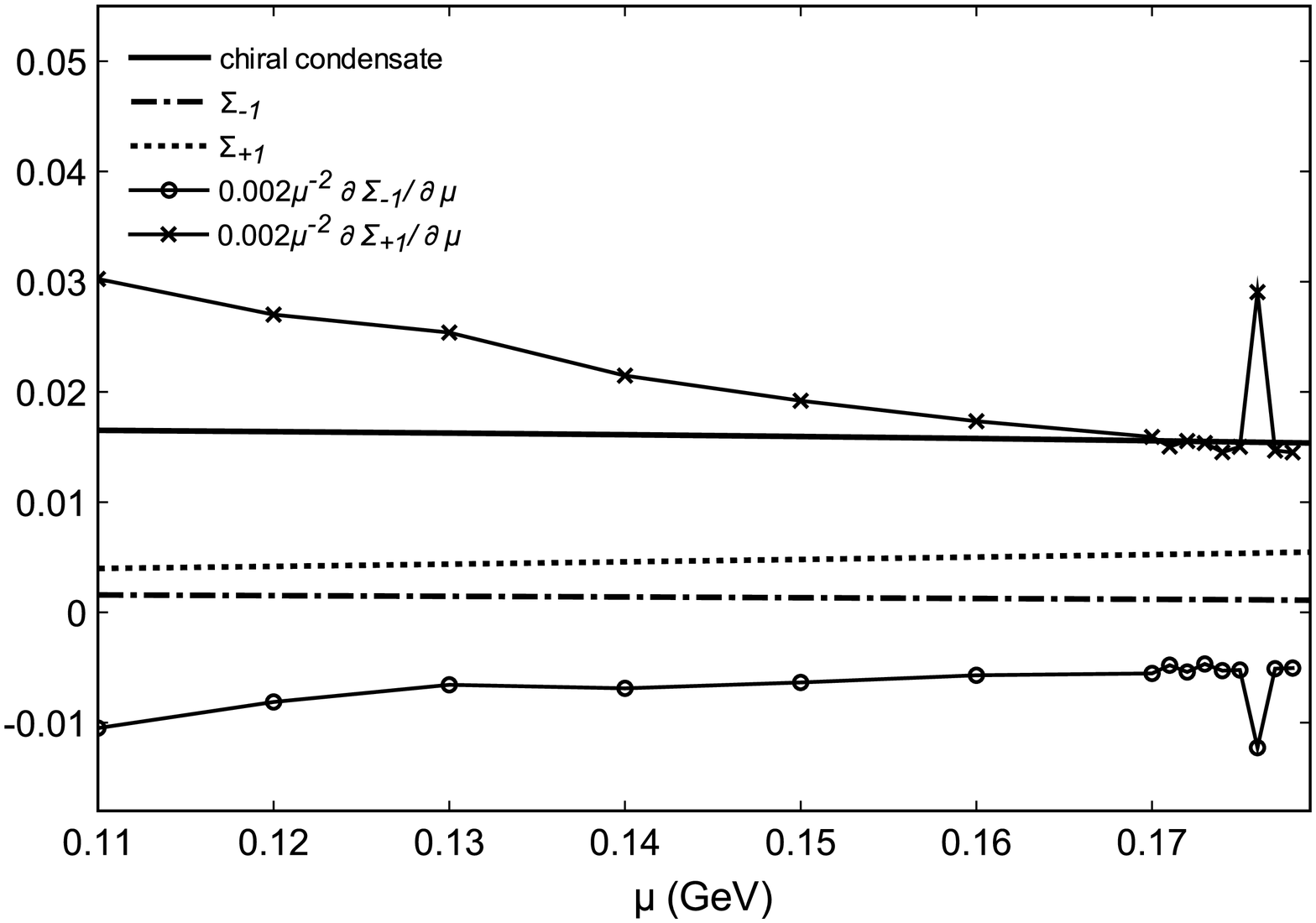}

\caption{The chemical potential dependence of the chiral condensate, the dressed Polyakov loop $\Sigma_{\pm 1}$, and $\mu^{-2}\partial\Sigma_{\pm 1}/\partial\mu$  at $T=150$ (MeV). The critical chemical potential obtained by the effective potential for the chiral phase transition is $127$ (MeV)}

\label{fig:pol2}

\end{center}

\end{minipage}

\end{tabular}

\end{figure}

\vspace{1em}
Our results are shown Figs\ \ref{fig:pol1} and \ref{fig:pol2}. The critical temperature for deconfinement phase transition is identical with the chiral phase transition. This result for the critical temperature agrees with \cite{rf:deco}. On the oher hand, the critical chemical potential is different even in the chiral limit. The difference is about 50 (MeV) at $T=150$ (MeV). (The critical chemical potential for the chiral phase transition critaical point is determined by the effective potential).

\vspace{1em}

\newpage

\vspace{1em}
\vspace{1em}

\vspace{1em}
\end{document}